# Simulation of a spatially correlated turbulent velocity field using biorthogonal decomposition

Pascal Hémon

*Department of Mechanics, LadHyX*

*Ecole Polytechnique-CNRS,*

*F-91128 Palaiseau, France*

*Tel. +33 1 69 33 39 33 – Fax +33 1 69 33 30 30*

*E-mail : pascal.hemon@ladhyx.polytechnique.fr*

Françoise Santi

*Department of Mathematics, Conservatoire National des Arts et Métiers,*

*292, rue Saint-Martin, F-75141 Paris, France*

**Abstract**

This paper presents a method for generating a turbulent velocity field that can be used as an input for the temporal simulation in wind excited structure problems. Temporal simulations become necessary when nonlinear behaviour, in the structure or in aeroelastic forces, must be accounted for. The main difficulty is then to reproduce correctly the statistical properties of the atmospheric turbulence, especially the spatial correlation. These properties constitute here the targets that the generated signal has to satisfy. We propose to use the biorthogonal decomposition technique which possesses interesting features to reach this objective, notably the space-time symmetry. Moreover, the convergence in energy is obtained rapidly with few terms of the decomposition, particularly in the low frequency range. Thus the method is found suitable for application to large civil engineering structures, such as bridges. Examples are provided for two different kinds of wind.

## 1.    Introduction

In the field of wind-excited civil engineering structures, temporal simulations are increasingly important, due to the large size of the structures as in modern suspended bridges [3]. Nowadays, the flutter problems are well known and the main difficulty arising in practice for the designers is to protect the structure from the turbulence effects. Particularly, a bridge under erection is delicate and sensitive to turbulent gusts.

As a consequence, there is a need for more accurate computations which can include the nonlinear behaviour of the structure, especially when cables vibrations must be taken into account. The classical techniques for solving the dynamical problem are generally based on spectral methods in which the nonlinear behaviours are difficult to introduce.

Temporal simulations do not present this default, and allow the direct coupling of the phenomena. Moreover, one can avoid the quadratic recombination of the eigenmodes response so that temporal simulations can provide better structural stress estimations.

To illustrate this, consider a bridge deck submitted to a mean wind velocity $\overline{U}$ associated to longitudinal and vertical turbulent components $u(t)$ and $w(t)$, as sketched in Figure 1. The resulting lift force acting on the deck is noted $F_z(t)$. This is the sum of the turbulent forces and the galloping force. Using the linear quasi-steady theory, it can be written as

$$F_z(t) = \frac{1}{2}\rho\, S\, \overline{U}^2 \left( \left( 1 + 2\frac{u(t)}{\overline{U}} \right) C_z + \frac{\partial C_z}{\partial \alpha}\frac{w(t)}{\overline{U}} - \frac{\partial C_z}{\partial \alpha}\frac{\dot{z}(t)}{\overline{U}} \right), \tag{1}$$

where $C_z$ is the lift coefficient of the deck, $S$ the reference surface, $\rho$ the air density, $\alpha$ the angle of attack and $\dot{z}(t)$ the vertical velocity of the deck motion.

In this expression, the turbulent velocity components appear explicitly in function of time. To calculate the response a long bridge deck, it is essential to account for the simultaneity and the size of the turbulent gusts along the bridge span.

Therefore the turbulent velocity components must satisfy the statistical wind properties, which depend on the local condition of the site. The objective of this paper is to present a method for generating such velocity field. The technique was shortly described in a recent paper [7]. First we will present the targets that we have chosen to simulate the signal. The method of construction is then presented and finally applied to realistic test cases.

## 2.    Wind characteristics

In the civil engineering community, the atmospheric turbulence is described with a small number of statistical parameters. We will deal in the following with the typical configuration shown in Figure 1, *i.e.* the longitudinal and vertical velocity components, which apply to an elongated horizontal structure placed on the *y* axis. These are (i) the mean velocity $\overline{U}(z)$, which may be a function of altitude *z*, (ii) the standard deviations of the velocity $\sigma_u$ (longitudinal) and $\sigma_w$ (vertical), (iii) the corresponding power spectral densities (PSD) $S_u(f)$ and $S_w(f)$ versus the frequency *f*, and (iv) the coherence functions in the lateral direction $\gamma_u^y(f)$ and $\gamma_w^y(f)$.

The chosen PSD functions are those proposed by von Kármán [3],

$$\frac{S_u(f)}{\sigma_u^2} = \frac{\dfrac{4\,\lambda_u^x}{\overline{U}(z)}}{\left(1+70.7\left(\dfrac{f\,\lambda_u^x}{\overline{U}(z)}\right)^2\right)^{5/6}}, \qquad \frac{S_w(f)}{\sigma_w^2} = \frac{\dfrac{4\lambda_w^x}{\overline{U}(z)}\left(1+188.4\left(\dfrac{2f\,\lambda_w^x}{\overline{U}(z)}\right)\right)^2}{\left(1+70.7\left(\dfrac{2f\,\lambda_w^x}{\overline{U}(z)}\right)^2\right)^{11/6}} \qquad (2)$$

where $\lambda_u^x$ and $\lambda_w^x$ are the longitudinal scales of $u$ and $w$ respectively.

The coherence functions are approximated by usual exponential functions

$$\gamma_u^y(f) = \exp\left[\frac{-C_u^y \left| y_i - y_j \right| f}{\overline{U}(z)}\right], \tag{3}$$

where the decay coefficient is $C_u^y$. The other velocity component $\gamma_w^y(f)$ is defined similarly with $C_w^y$. The cross-coherence function is expressed by combination of the single component functions:

$$\gamma_{uw}^{i,j}(f) = \sqrt{\gamma_u^{i,j}(f)\ \gamma_w^{i,j}(f)}. \tag{4}$$

## 3. Biorthogonal decomposition

### 3.1 General method

The biorthogonal decomposition (BOD) has been introduced by Aubry *et al.* [1] and the rigorous mathematical formulation can be found in that paper. The main idea is to carry out a deterministic decomposition of a space-time signal, *i.e.* the turbulent velocity field, by assuming its square-integrability only.

The BOD of a given signal $\boldsymbol{U}(\boldsymbol{x},t)$ function of space $\boldsymbol{x} \in \Re^3$ and time $t \in \Re$, with $\boldsymbol{U}(\boldsymbol{x},t) \in L^2(\boldsymbol{X} \times T)$, $\boldsymbol{X} \subset \Re^3$ and $T \subset \Re$, is formally written as

$$\boldsymbol{U}(\boldsymbol{x},t) = \sum_{k=1}^{\infty} \alpha_k\ \psi_k(t)\ \varphi_k(\boldsymbol{x}). \tag{5}$$

The BOD theorem proves that decomposition (6) exists, converges in norm and that

$$\alpha_1 \geq \alpha_2 \geq \ldots \geq 0, \quad \lim_{M \to \infty} \alpha_M = 0, \quad \langle \varphi_k, \varphi_l \rangle = \overline{\psi_k\ \psi_l} = \delta_{k,l}. \tag{6}$$

Aubry *et al.* have called *topos* the spatial modes $\varphi_k(\boldsymbol{x})$ with $\varphi_k \in L^2(\boldsymbol{X})$ and *chronos* the temporal modes $\psi_k(t)$ with $\psi_k \in L^2(\boldsymbol{T})$. They proved that the topos, associated to the set of the eigenvalues $\alpha_k^2 = \lambda_k$ are the eigenmodes of the spatial correlation operator

$$\boldsymbol{Sc}(\boldsymbol{x}, \boldsymbol{x}') = \int_T \boldsymbol{U}(\boldsymbol{x}, t) \boldsymbol{U}(\boldsymbol{x}', t) \, dt \,. \tag{7}$$

Simultaneously, the chronos associated to the same set of eigenvalues $\lambda_k$ are the eigenmodes of the temporal correlation operator

$$\boldsymbol{Tc}(t, t') = \int_X \boldsymbol{U}(\boldsymbol{x}, t) \boldsymbol{U}(\boldsymbol{x}, t') \, d\boldsymbol{x} \,. \tag{8}$$

What is remarkable is the fact that the eigenvalues $\alpha_k^2$ are common to topos and chronos: this was proved by using the symmetry property of the correlation operators [1]. This means that chronos and topos are intrinsically coupled because they have the same eigenvalue. However, it is possible to separate the information, spatial and temporal, by multiplying them by the weight factor $\sqrt{\alpha_k}$ .

The global energy of the signal is equal the sum of the eigenvalues:

$$\sum_{k=1}^{\infty} \alpha_k^2 = \text{Tr}(\boldsymbol{Sc}) = \text{Tr}(\boldsymbol{Tc}) \,. \tag{9}$$

The useful result in practice is the possibility to truncate decomposition (5) to *M* spatio-temporal structures.

It is important to recall here that the biorthogonal decomposition is deterministic and does not assume stationary and Gaussian signal, as the classical proper orthogonal decomposition (POD) or similar techniques do. Therefore the BOD can be used even with signals which the record is too short for standard analysis, as for instance it is commonly the case in climate observation [9].

### 3.2  Application to wind field generation

Turbulent velocity field have already been generated by various techniques. There exist a number of methods for generating a correlated turbulent velocity field as presented in the review by Guillin & Crémona and Di Paola [6, 4]. One of them is derived from the method proposed by Yamazaki & Shinozuka [9] for application in earthquake engineering. Their approach which is called *statistical preconditioning,* is based on the modal decomposition of the spatial covariance matrix and the temporal part of the signal is generated by using a Fourier decomposition. Another technique was proposed by Sakamoto and Ghanem [8] where the target is specified by the density functions of the process and the two point correlation functions. The spatial characteristics are recovered using a Karhunen-Loève expansion while the time characteristics are obtained through a polynomial chaos expansion.

Recently, Carassale & Solari [2] similarly used the direct proper orthogonal decomposition and a Fourier decomposition to generate a turbulent wind velocity field and to compute the wind loads acting on the eigenmodes of a structure.

These methods can be improved by exploiting the space-time symmetry of the BOD, as outlined in this paper. The new idea here is to build the velocity field as a BOD, which leads for the vertical component

$$w(y,t) = \sum_{m=1}^{M} \sqrt{\alpha_m^t} \; \psi_m(t) \; \sqrt{\alpha_m^y} \; \varphi_m(y) \tag{10}$$

where the chronos are associated with the set of eigenvalues $\left(\alpha_m^t\right)^2$ and the topos with $\left(\alpha_m^y\right)^2$. The main point is to find the topos and the chronos separately by solving the two corresponding eigenvalue problems. The principle is fundamentally different from the methods mentioned above because the basis functions arising in decomposition (10), i.e.

the topos and the chronos, are not chosen a priori but constructed in order to fit with the target properties.

The method of signal generation with BOD requires three main steps: (i) assembly of the two correlation matrices, spatial and temporal, (ii) resolution of the two related eigenvalues problems and (iii) generation of the velocity field using (10). It is also recommended a further step which consists in verification of the wind field properties by comparison with the targets.

The way the correlation matrices are assembled, which is given hereafter, can be subject to modifications and improvements. Hence we describe the technique we currently use, but any other techniques are possible at this stage.

The spatial correlation matrix is built starting from the PSD and coherence functions between nodes $i$ and $j$ as

$$\boldsymbol{Sc}_{i,j}^{w} = \sum_l \sqrt{S_{w_i}(f_l) S_{w_j}(f_l)} \; \gamma_w^y(f_l), \tag{11}$$

where $l$ refers to the frequency. The expressions are derived for the vertical velocity component, the longitudinal velocity component $u(y,t)$ being built using the same procedure.

For a 2D simulation, including both $u(y,t)$ and $w(y,t)$, the correlation matrix can be derived as in (11) but with the help of cross-coherence function, leading to

$$\boldsymbol{Sc}_{i,j}^{uw} = \sum_l \sqrt{\sqrt{S_{u_i}(f_l) S_{u_j}(f_l)} \sqrt{S_{w_i}(f_l) S_{w_j}(f_l)}} \; \gamma_{uw}^{i,j}(f_l). \tag{12}$$

In this case the 2D correlation matrix is written as

$$\boldsymbol{Sc} = \begin{pmatrix} \left[\boldsymbol{Sc}_{i,j}^{u}\right] & sym. \\ \left[\boldsymbol{Sc}_{i,j}^{uw}\right] & \left[\boldsymbol{Sc}_{i,j}^{w}\right] \end{pmatrix}, \tag{13}$$

and we see then that the size of the eigenvalue problem is multiplied by 4.

The temporal correlation matrix is given by

$$\boldsymbol{Tc}_{k,n}^{w} = \sum_i w_i(t_k)\, w_i(t_n) \tag{14}$$

where we assume that the individual signals at point $i$ and time $t_k$ are built with Fourier series as

$$w_i(t_k) = \sum_l \sqrt{2\, S_{w_i}(f_l)}\ \cos\!\left(2\pi\, f_l\, t_k + \phi_{i,l}\right). \tag{15}$$

The phase angles $\phi_{i,l}$ are randomly uniformly distributed in $[0, 2\pi]$. Note that the present choice for the temporal correlation matrix is arbitrary and might be improved, but this does not influence the overall method presented here.

In case of a 2D simulation, the cross-correlation needs to be taken, leading to a matrix of the form

$$\boldsymbol{Tc} = \begin{pmatrix} \left[\boldsymbol{Tc}_{k,n}^{u}\right] & sym. \\ \left[\boldsymbol{Tc}_{k,n}^{uw}\right] & \left[\boldsymbol{Tc}_{k,n}^{w}\right] \end{pmatrix}. \tag{16}$$

From the relations given above, it must be noticed then that the number of time steps and the number of nodes on the structure should be equal, which might be a constraint in practice.

## 4.    Test cases

### 4.1    Presentation

The illustration of the above method is performed on two kinds of atmospheric wind. Case A is a sea wind and corresponds to the Saint-Nazaire bridge in France, at an altitude of 65 m above sea level ($\overline{U}$ =40.9 m/s). Case B is a mountain wind, corresponding to the Millau bridge at 270 m altitude ($\overline{U}$ =36.5 m/s). Parameters are given in Table 1.

Note that we deal here with a horizontal bridge deck, where the turbulence level is constant along its span. Then the turbulence is homogeneous in these test cases, although the method can be applied to vertical structures for which the incoming turbulence depends on altitude. As underlined recently by Farge et al. [5], the orthogonal decompositions based on correlation operators degenerate in Fourier decomposition when it is applied to homogeneous turbulence. The proposed method based on BOD is therefore a more general technique which Fourier decomposition is a particular case.

### 4.2   Results

The following results have been obtained with 256 time steps and nodes, a sampling frequency of 6 Hz and a frequency band of 0.075-3 Hz. It is important to recall here that the time discretization, and subsequently the space discretization, has to be chosen properly for the problem: especially the Shannon theorem must be respected, and the signal duration must be sufficient for representing the lowest frequency. In the present test, the generated signal is 42.6 s long, which represent only 3.2 periods at the lowest frequency of 0.075 Hz.

Note that in the Fourier series (15), the frequency band is discretized following a logarithmic law, in order to correctly represent the lowest frequencies. The deck spanwise length is 350 m.

The BOD is restricted to a number $M$ of spatio-temporal structures. It is therefore important to check the convergence of the method and this can be done with the energy defined in equation (9). But the critical point is the fact that the signal is generated with the help of topos and chronos that are computed separately, as in decomposition (10).

This assumes that the convergence of the topos and the chronos is obtained simultaneously.

The Figure 2 presents the cumulated energy of case A for the two velocity components $u$ and $w$. About 90% of energy is reproduced in the signal with the 40 first terms of the BOD and 95% with 80 terms. Moreover, topos and chronos converge at the same rate, a small difference appearing at higher orders.

Samples of velocities are given in Figure 3 versus time. The calculated standard deviations are a little lower than the target one, due to the truncation to 80 terms, which produces an energy deficit of 2.5 %.

Comparison of resulting spectra with the target function, given in Figure 4, shows the good agreement. These spectra are directly computed with the Fourier transform of single temporal signals which the time resolution and length are mentioned previously. The noise and the lower agreement at low frequency is a direct consequence of this choice (256 points sampled at 6 Hz). However, even with such constraints the simulated signals follow the target function. When the number of terms $M$ in the BOD is decreased from 80 to 40, see Figure 5, the global level of the PSD decreases due to the lower level of the energy simulated. Moreover, we observe simultaneously that decreasing the number of terms acts as a low pass filter (at about 0.7 Hz for $M$=40).

Similarly the correlation functions compare well in Figure 6. Moreover the cross-correlation between longitudinal and vertical component, usually difficult to match, shows a good agreement in Figure 7.

5.    **Conclusion**

Generation of a spatially correlated velocity field can be easily performed by BOD within an elegant formulation. The method takes advantage of the space-time symmetry of the BOD which requires only the square integrability of the velocity field. The number of spatio-temporal structures taken in the decomposition is the parameter that fixes the RMS level and the frequency band. The energy criterion is easy to check and it is recommended to truncate the decomposition when 95% of the energy is recovered.

The spatial correlation of the generated signal is found in good agreement with the specified targets.

Extension of the technique to other applications is easy, for instance boundary input conditions for LES calculations.

## 6.    References

[1]    N. Aubry, R. Guyonnet, R. Lima, Spatiotemporal analysis of complex signals: Theory and applications. J. of Statistical Physics 64 (1991) 683-739.

[2]    L. Carassale, G. Solari, Wind modes for structural dynamics: a continuous approach. Probabilistic Engineering Mechanics 17 (2002) 157-166.

[3]    C. Crémona, J.-C. Foucriat (eds), Comportement au vent des ponts. Presses de l'Ecole Nationale des Ponts et Chaussées, (2002), Paris, France.

[4]    M. Di Paola, Digital simulation of wind field velocity. J. of Wind Eng. and Indus. Aerod. 74-76 (1998) 91-109.

[5]    M. Farge, K. Schneider, G. Pellegrino, A.A. Wray, R.S. Rogallo, Coherent vortex extraction in 3-D homogeneous turbulence: comparison between CVS-wavelet and POD-Fourier decompositions. Physics of Fluids 15(10) (2003) 2886-2896.

[6]    A. Guillin, C. Crémona, Développement d'algorithmes de simulation de champs

de vitesse de vent. Report of Laboratoire Central des Ponts et Chaussées, section

Ouvrages d'Art, OA27 (1997), Paris, France.

[7]    P. Hémon, F. Santi, Applications of biorthogonal decompositions in fluid-

structure interactions. J. of Fluids and Structures 17(8) (2003) 1123-1143.

[8]    S. Sakamoto, R. Ghanem, Simulation of multi-dimensional non-gaussian non-

stationnary random fields. Prob. Eng. Mechanics 17 (2002) 167-176.

[9]    R. Vautard, P. Yiou, M. Ghil, Singular-spectrum analysis: A toolkit for short,

noisy chaotic signals. Physica D 58 (1992) 95-126.

[10]   F. Yamazaki, M. Shinozuka, Simulation of Stochastic Fields by Statistical

Preconditioning. ASCE J. of Engineering Mechanics 116 (1990) 268-287.

FIGURE CAPTIONS

Figure 1: Typical configuration

Figure 2: Convergence, $u$ (left) and $w$ (right) components, Case A

Figure 3: Samples of velocities, $u$ (left) and $w$ (right) components, Case A

Figure 4: Comparison of PSD, $u$ (left) and $w$ (right) components, Case B, $M$=80

Figure 5: Comparison of PSD, $u$ (left) and $w$ (right) components, Case B, $M$=40

Figure 6: Comparison of spatial correlation, $u$ (left) and $w$ (right) components, Case B

Figure 7: Comparison of spatial cross-correlation $u$-$w$, Case B (left), Case A (right)

| component | Sea wind, case A | | | Mountain wind, case B | | |
|---|---|---|---|---|---|---|
| | $\sigma/\overline{U}$ | $\lambda^y$ (m) | $C^y$ | $\sigma/\overline{U}$ | $\lambda^y$ (m) | $C^y$ |
| $u$ | 0.09 | 85 | 11 | 0.16 | 90 | 12 |
| $w$ | 0.05 | 35 | 12 | 0.12 | 30 | 9 |

Table 1: Parameters simulated winds

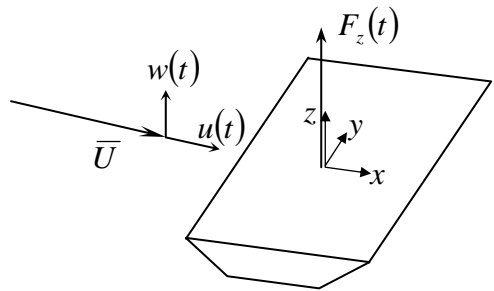

Figure 1: Typical configuration

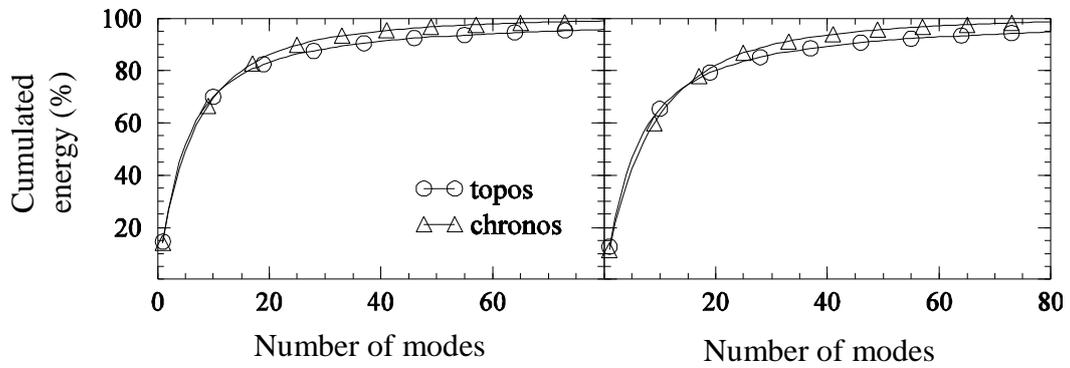

Figure 2: Convergence, $u$ (left) and $w$ (right) components, Case A

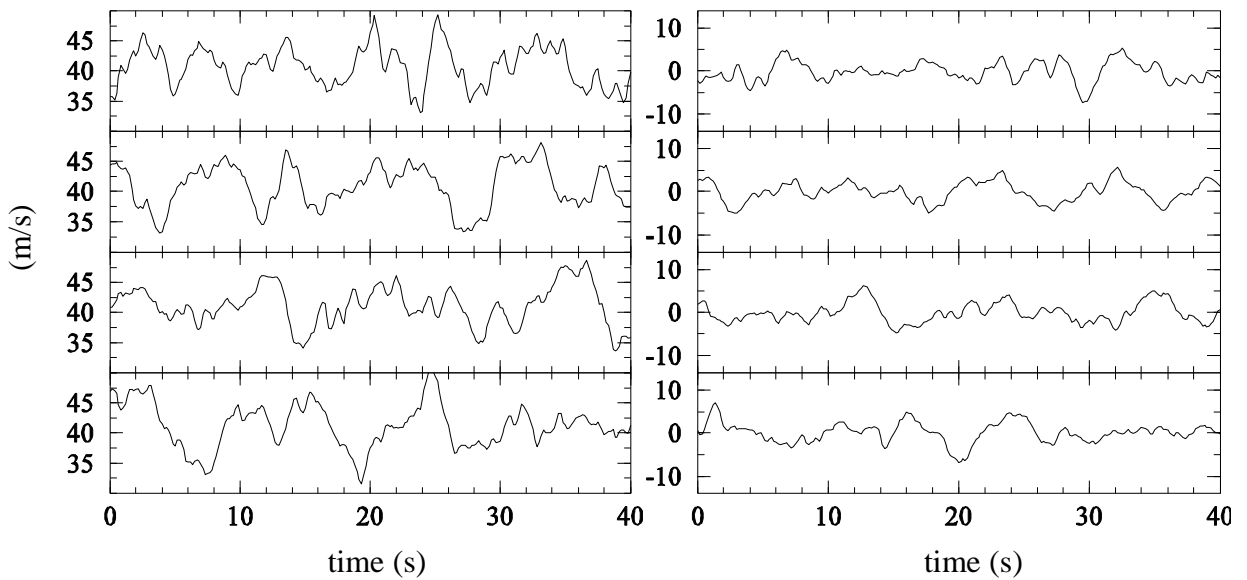

Figure 3: Samples of velocities, $u$ (left) and $w$ (right) components, Case A

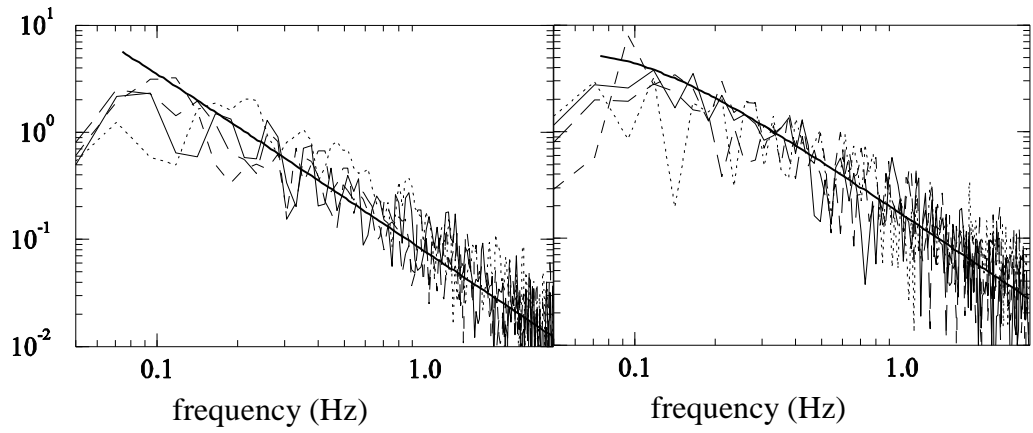

Figure 4: Comparison of PSD, *u* (left) and *w* (right) components, Case B, *M*=80

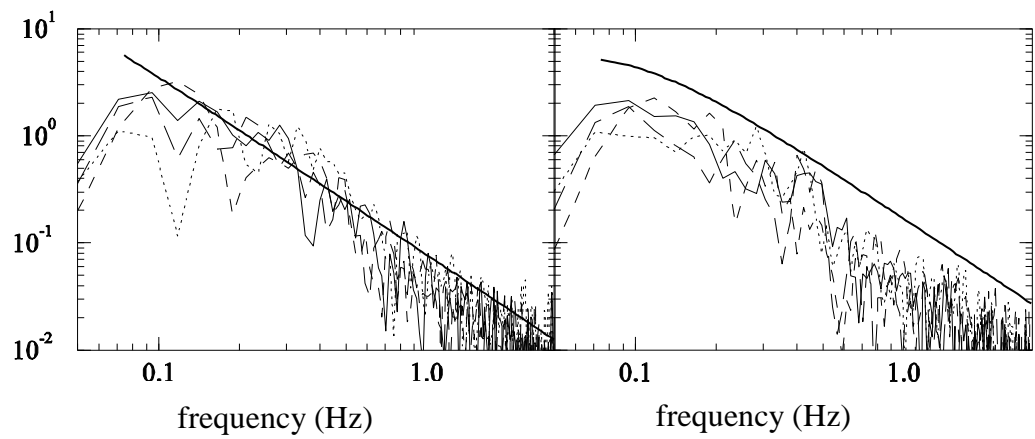

Figure 5: Comparison of PSD, *u* (left) and *w* (right) components, Case B, *M*=40

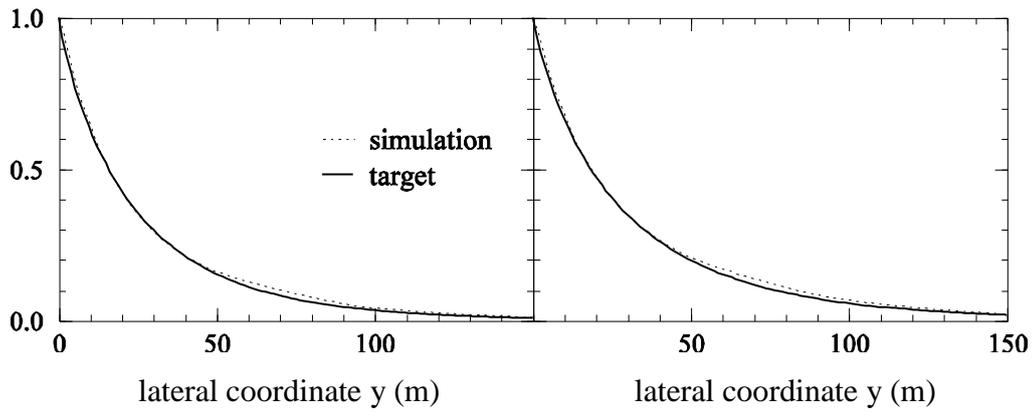

Figure 6: Comparison of spatial correlation, *u* (left) and *w* (right) components, Case B

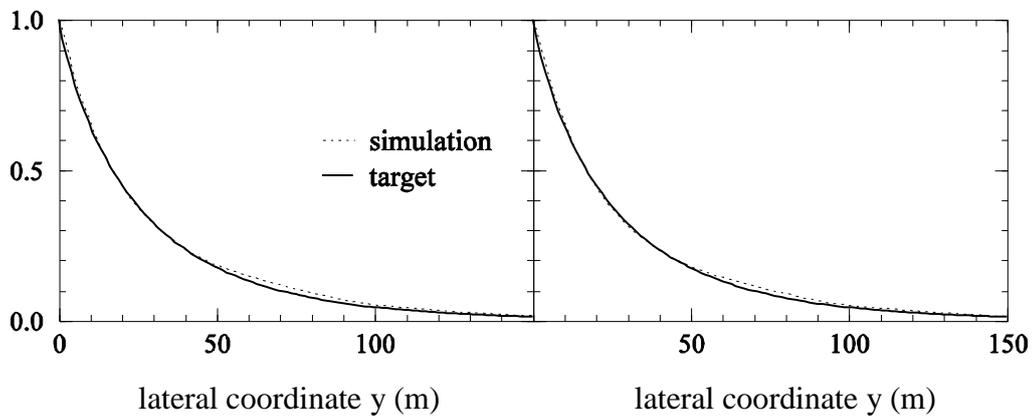

Figure 7: Comparison of spatial cross-correlation *u-w*, Case B (left), Case A (right)